\documentclass[twocolumn]{aastex631}

\newcommand{\HI}{H\,{\sc i}}

\shorttitle{AASTeX v6.3.1 Sample article}
\shortauthors{Brown et al.}
\graphicspath{{./}{figures/}}

\begin{document}

\title{VERTICO VII: Environmental quenching caused by suppression of molecular gas content and star formation efficiency in Virgo Cluster galaxies}

\correspondingauthor{Toby Brown}
\email{tobiashenrybrown@gmail.com}

\author[0000-0003-1845-0934]{Toby Brown}\affiliation{Herzberg Astronomy and Astrophysics Research Centre, National Research Council of Canada, 5071 West Saanich Rd, Victoria, BC, V9E 2E7, Canada}

\author{Ian D. Roberts}\affiliation{Leiden Observatory, Leiden University, PO Box 9513, 2300 RA Leiden, The Netherlands}

\author{Mallory Thorp}\affiliation{Department of Physics \& Astronomy, University of Victoria, Finnerty Road, Victoria, BC, V8P 1A1, Canada}

\author{Sara L. Ellison}\affiliation{Department of Physics \& Astronomy, University of Victoria, Finnerty Road, Victoria, BC,   V8P 1A1, Canada}

\author[0000-0001-7732-5338]{Nikki Zabel}\affiliation{Department of Astronomy, University of Cape Town, Private Bag X3, Rondebosch 7701, South Africa}

\author{Christine D. Wilson}\affiliation{Department of Physics \& Astronomy, McMaster University, 1280 Main Street W, Hamilton, ON, L8S 4M1, Canada}

\author{Yannick M.~Bah\'{e}}\affiliation{Leiden Observatory, Leiden University, PO Box 9513, 2300 RA Leiden, The Netherlands}

\author{Dhruv Bisaria}\affiliation{Department of Physics, Engineering Physics and Astronomy, Queen’s University, Kingston, ON K7L 3N6, Canada}

\author{Alberto D. Bolatto}\affiliation{Department of Astronomy, University of Maryland, College Park, MD, 20742, USA)}\affiliation{Visiting Scholar at Flatiron Institute, Center for Computational Astrophysics (NY 10010, USA) }

\author{Alessandro Boselli}\affiliation{Aix-Marseille Universit\'{e}, CNRS, CNES, LAM, Marseille, France}

\author{Aeree Chung}\affiliation{Department of Astronomy, Yonsei University, 50 Yonsei-ro, Seodaemun-gu, Seoul 03722, South Korea}

\author{Luca Cortese}\affiliation{International Centre for Radio Astronomy Research, The University of Western Australia, 35 Stirling Hwy, 6009 Crawley, WA, Australia }\affiliation{ARC Centre of Excellence for All Sky Astrophysics in 3 Dimensions (ASTRO 3D), Australia}

\author{Barbara Catinella}\affiliation{International Centre for Radio Astronomy Research, The University of Western Australia, 35 Stirling Hwy, 6009 Crawley, WA, Australia }\affiliation{ARC Centre of Excellence for All Sky Astrophysics in 3 Dimensions (ASTRO 3D), Australia}

\author[0000-0003-4932-9379]{Timothy A. Davis}\affiliation{Cardiff Hub for Astrophysics Research \&\ Technology, School of Physics \&\ Astronomy, Cardiff University, Queens Buildings, Cardiff, CF24 3AA, UK}

\author{Mar\'{i}a J. Jim\'{e}nez-Donaire}\affiliation{Observatorio Astronómico Nacional (IGN), C/Alfonso XII, 3, E-28014 Madrid, Spain}\affiliation{Centro de Desarrollos Tecnológicos, Observatorio de Yebes (IGN), 19141 Yebes, Guadalajara, Spain}

\author{Claudia D.P. Lagos}\affiliation{International Centre for Radio Astronomy Research, The University of Western Australia, 35 Stirling Hwy, 6009 Crawley, WA, Australia }\affiliation{ARC Centre of Excellence for All Sky Astrophysics in 3 Dimensions (ASTRO 3D), Australia}

\author{Bumhyun Lee}\affiliation{Korea Astronomy and Space Science Institute, 776 Daedeokdae-ro, Daejeon 34055, Republic of Korea}

\author{Laura C. Parker}\affiliation{Department of Physics \& Astronomy, McMaster University, 1280 Main Street W, Hamilton, ON, L8S 4M1, Canada}

\author{Rory Smith}\affiliation{Departamento de F\'{i}sica, Universidad T\'{e}cnica Federico Santa Mar\'{i}a, Avenida Vicuña Mackenna 3939, San Joaqu\'{i}n, Santiago de Chile}

\author{Kristine Spekkens}\affiliation{Royal Military College of Canada, PO Box 17000, Station Forces, Kingston, ON, Canada K7K 7B4}

\author{Adam R.H. Stevens}\affiliation{International Centre for Radio Astronomy Research, The University of Western Australia, 35 Stirling Hwy, 6009 Crawley, WA, Australia }\affiliation{ARC Centre of Excellence for All Sky Astrophysics in 3 Dimensions (ASTRO 3D), Australia}

\author{Vicente Villanueva}\affiliation{Department of Astronomy, University of Maryland, College Park, MD 20742, USA}

\author{Adam B. Watts}\affiliation{International Centre for Radio Astronomy Research, The University of Western Australia, 35 Stirling Hwy, 6009 Crawley, WA, Australia }




\begin{abstract}
We study how environment regulates the star formation cycle of 33 Virgo Cluster satellite galaxies on $720$ parsec scales. 
We present the first resolved star-forming main sequence for cluster galaxies, dividing the sample based on their global \HI~ properties and comparing to a control sample of field galaxies. \HI-poor cluster galaxies have reduced star formation rate (SFR) surface densities with respect to both \HI-normal cluster and field galaxies ($\sim 0.5~$ dex), suggesting that mechanisms regulating the global \HI~content are responsible for quenching local star formation.
We demonstrate that the observed quenching in \HI-poor galaxies is caused by environmental processes such as ram pressure stripping (RPS) simultaneously reducing molecular gas surface density {\em and} star formation efficiency (SFE), compared to regions in \HI-normal systems (by 0.38 and 0.22 dex, respectively).
We observe systematically elevated SFRs that are driven by increased molecular gas surface densities at fixed stellar mass surface density in the outskirts of early-stage RPS galaxies, while SFE remains unchanged with respect to the field sample.
We quantify how RPS and starvation affect the star formation cycle of inner and outer galaxy discs as they are processed by the cluster. We show both are effective quenching mechanisms with the key difference being that RPS acts upon the galaxy outskirts while starvation regulates the star formation cycle throughout disc, including within the truncation radius. For both processes, the quenching is caused by a simultaneous reduction in molecular gas surface densities and SFE at fixed stellar mass surface density.
\end{abstract}



\section{Introduction} \label{sec:intro}

Understanding the gas-star formation cycle in galaxy clusters has been an active area of research for more than four decades \citep[see reviews by][and references therein]{Haynes1984, Boselli2006, Boselli2014review, Cortese2021, Boselli2022, Alberts2022}. It is now clear that there are a large number of environmental processes that affect how star formation proceeds in cluster satellite galaxies, often concurrently (e.g., starvation -- \citealt{Larson1980}, tidal stripping -- \citealt{Moore1999}, thermal evaporation -- \citep{Cowie1977}, gravitational interaction -- \citealt{Moore1996}). A common attribute of these mechanisms is that they primarily exert their influence on galaxy evolution by perturbing the interstellar medium (ISM). 

Atomic hydrogen gas (\HI) is usually the most massive and extended component of the ISM, extending well beyond the galaxy stellar radius \citep{Cayatte1994}, making it an excellent tracer of the environmental influence on galaxies. As such, and given the efficiency with which \HI~reservoirs can be mapped, a large number of studies have confirmed the systematic depletion of \HI~reservoirs by external mechanisms \citep[e.g.,][]{Giovanelli1985, Solanes2001, Gavazzi2008, Chung2009, Cortese2011, Serra2012, Catinella2013, Brown2017, Stevens2017, Stevens2019, Healy2020}. 

Although less well studied than \HI, there is a growing body of work revealing the role of environment in perturbing galaxies' molecular gas content, as the direct fuel for star formation \citep{Boselli2002, Fumagalli2009, Wilson2009, Corbelli2012, Boselli2014, Lee2017, Mok2016,Nehlig2016, Moretti2018, Zabel2019, Jachym2019, Cramer2020, Moretti2020a, Moretti2020b, Brown2021, Stevens2021, Morokuma-Matsui2022, Roberts2022, Lee2022}. 

Recent years have seen a particular focus on the role of ram pressure stripping in both quenching and inducing star formation in cluster galaxies. Ram pressure stripping occurs in galaxies moving at high velocity through the intra-cluster medium (ICM), where pressure exerted on the ISM is strong enough to directly remove gas from the disc. First proposed by \citet{Gunn1972}, there are now many examples of this mechanism affecting how galaxies evolve in dense environments \citep[e.g.,][]{Abadi1999,Hester2006,Fumagalli2014,Poggianti2017,Fossati2016,Roberts2021}. Indeed, this topic has been the subject of two recent review papers, \citet{Cortese2021} and \citet{Boselli2022}, and we refer the interested reader to those works for a comprehensive overview.

There is clear evidence that ram pressure stripping is capable of quenching satellite galaxies, primarily through removal of molecular gas and/or the suppression of star formation efficiency (SFE) throughout the disc \citep[e.g.,][]{Mok2017, Moretti2018, Zabel2020, Zabel2022, Villanueva2022}. However, there are also a growing number of studies showing elevated star formation in galaxies that are actively undergoing stripping, suggesting that the pressure at the ICM--ISM interface acts to increase SFE or aid the conversion of \HI~into molecular gas \citep[e.g.,][]{Ebeling2014, Nehlig2016, Vulcani2018, Lizee2021, Moretti2020a, Moretti2020b, Roberts2022}.

The lack of widespread rejuvenation of star formation in cluster galaxies means that starvation -- the cessation of gas supply to the ISM -- must also be an important regulatory mechanism in this environment. Indeed, the prevalence of starvation is also inferred by studies of slow quenching timescales in cluster galaxies \citep[e.g.,][]{McGee2009, Haines2015, Paccagnella2016}. For cluster galaxies, especially those that have undertaken one pericenter passage or more, it is highly likely that both starvation and direct gas stripping are playing a major role in the gas-star formation cycle. For such systems, we can therefore more clearly define the term starvation to mean the lack of gas accretion after parts of the disc have been stripped. Thus, the key question is not {\em if} but {\em how}, {\em where}, and {\em when} do ram pressure stripping and starvation affect the star formation cycle? Answering this requires establishing if the remaining gas reservoirs of stripped galaxies are affected by other environmental processes such as starvation, or if they proceed to form stars as though they were in the field. In other words, does the starvation mechanism itself break the star formation cycle, or does it simply stop it from starting again once it has been disrupted by another process?

The combined literature has led to a general consensus that environmental processes play an important role in regulating star formation in dense environments. However, the exact nature of different galaxy quenching mechanisms, and the extent to which they influence molecular gas content and SFE are questions that are still unanswered.

It is now common to use a measurement of \HI~deficiency as a proxy for environmental influence. This parameter quantifies the \HI~content with respect to the typical content of a comparable field galaxy. Most commonly, field control samples are constructed to match the target galaxy in either the size or mass of the stellar disc \citep{HaynesGiovanelli1984, Chung2009, Boselli2014, Zabel2022}. There are a number of studies demonstrating the efficacy of \HI~deficiency as a tracer of environmental influence, both for cluster galaxies in general \citep[e.g.,][]{Boselli2009, Li2020} and the specific galaxies studied in this work (e.g., \citealt[][]{Yoon2017, Zabel2022, Watts2023}. For an extensive discussion of \HI~deficiency, including the respective merits of the various definitions, we refer the reader to Section 3 of \citet{Cortese2021}.

In nearby field galaxies, studying the form and scatter of the spatially resolved star-forming main sequence (rSFMS), which relates the stellar mass surface density, $\Sigma_\star$, with star formation rate (SFR) surface density, $\Sigma_\mathrm{SFR}$, is an increasingly common approach to characterize the local star formation cycle. To first order, how regions evolve along and deviate from this relationship is determined by the processes that regulate the conversion of gas into stars \citep{Sanchez2013, Cano-Diaz2016, Hsieh2017, Lin2019, Enia2020, Ellison2020, Sanchez2020, Ellison2021, Baker2022, Bluck2022}. Recent works investigating the rSFMS in field galaxies on kiloparsec spatial scales present a picture where star formation is regulated locally (e.g., feedback, self-regulation) while quenching is more closely related to global mechanisms that affect the total gas reservoir \citep{Bluck2020, Ellison2020, Sanchez2020}. Thus, considering the cluster rSFMS is important for establishing how environmental mechanisms which affect the global \HI~content (as measured by \HI-deficiency) are capable of either inducing or quenching local star formation activity. Studying the resolved relation rather than the global SFR-stellar mass relationship also allows one to more readily identify the signatures of environmental quenching in different parts of the galaxy disc (e.g., inner disc versus outskirts, leading versus trailing edge, etc.).

Since cold molecular gas is the raw fuel for star formation \citep[e.g.,][]{Bigiel2008}, the relation between $\Sigma_\star$ and $\Sigma_\mathrm{SFR}$ (i.e., the rSFMS) can be understood as a consequence of the more fundamental connections between these two properties and molecular gas surface density, $\Sigma_\mathrm{mol}$ \citep{Lin2019, Pessa2021, Sanchez2021}.

The connection between the SFR and molecular gas surface densities is commonly known as the resolved Kennicutt-Schmidt relation (rKSR) and has been studied extensively in nearby field galaxies \citep[e.g.,][]{Bigiel2008, Leroy2013, Roychowdhury2015, Morselli2020}. Recent resolved studies of molecular gas in cluster galaxies have begun to explore the subkiloparsec rKSR in this regime for statistically significant samples \citep{Zabel2020, Jimenez-Donaire2022}. These studies show that, despite the large galaxy-to-galaxy variation, the cluster rKSR largely follows the same form as in field galaxies. However, \citet{Jimenez-Donaire2022} also demonstrate that environmental mechanisms that affect the galaxy \HI~content reduce the SFE of the molecular gas.

Although less well studied than the rKSR, a growing number of works have established the correlation between molecular gas and stellar surface densities -- known as the resolved molecular gas main sequence (rMGMS) -- which governs the amount of gas available for star formation \citep{Lin2019, Pessa2021, Ellison2021, Morselli2020}. The presence of the rMGMS in cluster members was recently established by \citet{Watts2023} who show that the individual cluster galaxy rMGMS gradually moves below the field relation as \HI~deficiency increases.

Recent field galaxy studies have analyzed connections between the rSFMS, rMGMS, and rKSR simultaneously for several different samples \citep[e.g.,][]{Lin2019, Ellison2020, Sanchez2021, Pessa2021, Baker2022}. \citet{Ellison2020} and \citet{Baker2022} demonstrate that the rSFMS in field galaxies is driven primarily by the form and scatter of the rKSR (i.e., SFE). A secondary but strong correlation with the rMGMS (i.e., gas content) is likely a consequence of the rKSR's slightly sub-linear nature in log--log space \citep{Ellison2021, Sanchez2021, Jimenez-Donaire2022}. Since the rSFMS is fundamentally a projection of the rMGMS and rKSR, it is critical to interpret this relation in the context of the other two. In taking this approach, the aim of this work is to provide a holistic view of how different environmental mechanisms regulate star formation activity in the VERTICO sample.

The paper is structured as follows: Section \ref{sec:data} describes the data and sample selection. We present evidence for the influence of environmental processes on the molecular gas-star formation cycle of Virgo galaxies in Section \ref{sec:results}. Section \ref{sec:discussion} discusses these results and our interpretation in the context of previous work. Lastly, we summarize our findings and briefly highlight areas for future research in Section \ref{sec:discussion}. Throughout this analysis, we assume a constant distance of 16.5 Mpc to all Virgo galaxies based upon the Virgo Cluster distance found by \citet{Mei2007}. SFRs and stellar masses are derived assuming a Chabrier initial mass function \citep{Chabrier2003}.

\section{Data}
\label{sec:data}

The data in this paper are drawn from two surveys of nearby star-forming galaxies. Our primary analysis sample of cluster galaxies is selected from the Virgo Environment Traced in CO survey \citep[VERTICO;][]{Brown2021} while our field control sample is selected from the Heterodyne Receiver Array CO Line Extra-galactic Survey \citep[HERACLES;][]{Leroy2009}. This section describes the data from each.

\subsection{The VERTICO Survey}
\label{sec:vertico}

VERTICO is a completed Atacama Large Millimeter/submillimeter Array (ALMA) Cycle 7 Large Program that uses ALMA's Morita array to map the distribution and kinematics of CO(2--1) across 51 Virgo Cluster galaxies on subkiloparsec scales. Here we briefly describe the VERTICO sample, CO(2--1) data products and derivation of molecular gas surface densities. For more details, we refer the reader to \citet{Brown2021}. From here on, we use ``CO'' to explicitly refer to the CO(2--1) transition. Other transitions are noted accordingly.

All resolved data used in this work are smoothed to the smallest common beam diameter of $9\arcsec$ with a pixel scale of $4\arcsec$. The $9\arcsec$ beam corresponds to a physical size of $720~$pc at the distance of Virgo \citep[$16.5~$Mpc, $1\arcsec \approx 80$~pc;][]{Mei2007}. Matching the beam size across all galaxies ensures that each resolution element covers approximately the same physical area of the disc, and the approximate Nyquist sampling with the pixel scale reduces the effect of beam oversampling contributing to trends in our analysis. Given the consistent physical scale of each pixel, we use the words `pixel' and `region' interchangeably to improve readability.

The molecular gas surface density maps in $\mathrm{M_\odot ~ pc^{-2}}$ are derived assuming a constant CO(1--0) conversion factor of $\alpha_{\rm CO} = 4.35 ~ {\rm M_\odot ~ pc^{-2} ~ (K~km~s^{-1})^{-1}}$ and CO(1--0)-to-CO(2--1) line luminosity ratio ($R_{21}$) of 0.8 \citep{Brown2021}. These maps include the $36\%$ contribution of helium in addition to H$_2$ and are corrected for projection effects using the optical $r$-band inclination.

From the VERTICO sample of 49 galaxies detected in CO, we selected all galaxies with an optical $r$-band inclination $\leq 80^{\circ}$ and pixels with a signal-to-noise ratio $\geq 2$. Further, we remove the three lowest stellar mass galaxies in our sample (NGC 4299, NGC 4532, NGC 4561) since these galaxies have low metallicity and their molecular gas component is likely underestimated by the constant $\alpha_{\rm CO}$ used in this paper. We also exclude NGC 4321 from our sample as the resolving beam of the ACA CO data is $\approx 10.5\arcsec$. This selection yields a final sample of 15,401 pixels drawn from 33 galaxies covering a range in projected cluster-centric distances between $\sim0.2\,r_{200}$ and $\sim2.5\,r_{200}$ \citep[$r_{200} = 1.55~$Mpc;][]{McLaughlin1999, Brown2021}.

\subsection{HERACLES}
\label{sec:heracles}

HERACLES is a survey of CO(2--1) emission in 48 nearby field galaxies ( $2 < D \, / \, {\rm Mpc} < 25$) spanning a comparable range in global stellar mass and sSFR to VERTICO ($10^{8.5} < {\rm M_\star \, / \, M_\odot} < 10^{11}$ and $10^{-11.5} < {\rm sSFR \, / \, yr^{-1}} < 10^{-9.2}$, respectively). The survey's public data cubes have an angular beam diameter of $13\arcsec$ in 5~km~s$^{-1}$ wide channels. We refer the reader to \citet{Leroy2009} for further details on the survey design and data products.

The HERACLES targets span a range of distances so the physical scales probed by the data's common $13\arcsec$ beam varies considerably. Therefore, for the comparison to VERTICO, we select the 10 closest galaxies ($2.9 \leq D/\mathrm{Mpc} \leq 10.6$) and convolve the cubes with a Gaussian kernel so that the final beam matches the $720~$pc physical resolution of the VERTICO $9\arcsec$ beam. This ensures we are probing the same physical scales with our comparison. We also smooth the data to the same $10~\mathrm{km~s^{-1}}$ velocity resolution as the final VERTICO data cubes. Science ready data products such as flux, gas surface density, and signal-to-noise maps are produced using an identical procedure to the VERTICO data products described in \citet{Brown2021}. For this work we apply the selection criteria described in Section \ref{sec:vertico}, yielding a final field control sample of 10,817 pixels from drawn from 10 HERACLES survey galaxies.

\subsection{Stellar mass and SFR surface densities}

This section briefly describes the production of $9\arcsec$~SFR and stellar mass surface density maps for both the VERTICO and HERACLES galaxies. A fuller description of this procedure can be found in \citet{Jimenez-Donaire2022} and \citet{Villanueva2022} for the SFR and stellar mass surface density maps, respectively.

All maps are constructed using GALEX and WISE photometry following the procedure laid out in \citet{Leroy2019}.
The final 9\arcsec resolutions match the beam size of the VERTICO CO products.  
SFR surface density maps are constructed from a combination of GALEX near-UV (NUV) and WISE3 photometry at 9\arcsec. We are required to use WISE3 as our obscured tracer at 9\arcsec~resolution as the WISE4 beam is too coarse.   
Stellar mass surface density maps are derived from WISE1 photometry. We determine the local mass-to-light ratio (at $3.4~\mathrm{\mu m}$) using the WISE3-to-WISE1 color as an ‘sSFR-like’ proxy and following the calibrations in \citet{Leroy2019}. The WISE1 images are then combined with the derived mass-to-light ratios to produce stellar mass surface density maps. As with the molecular gas, the stellar and SFR surface densities are corrected for inclination effects by a factor of cos($i$).

\begin{figure*}
    \centering
    \includegraphics{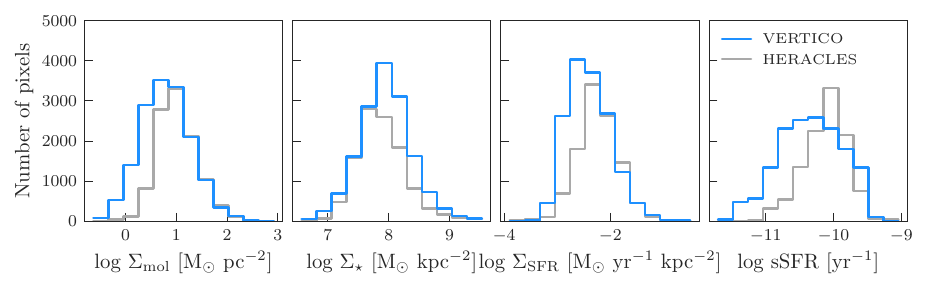}
    \caption{Comparison of resolved physical property distributions for the VERTICO cluster (blue) and HERACLES field (gray) samples. From left to right, molecular gas surface density ($\Sigma_\mathrm{mol}$), stellar surface density ($\Sigma_\star$), SFR surface density ($\Sigma_\mathrm{SFR}$), and specific SFR (sSFR). All surface density properties are corrected for projection effects using the optical inclination. As the cluster sample, regions in VERTICO galaxies are skewed to lower molecular gas and SFR densities.} 
    \label{fig:distributions}
\end{figure*}

Figure \ref{fig:distributions} shows the distributions of resolved physical properties for the 15,401 pixels in the VERTICO analysis sample and 10,817 pixels in the HERACLES field control sample. From left to right, the properties shown are the molecular gas surface density ($\Sigma_\mathrm{mol}$), stellar surface density ($\Sigma_\star$), SFR surface density ($\Sigma_\mathrm{SFR}$), and specific SFR (sSFR $= \Sigma_\mathrm{SFR} / \Sigma_\star$). The distribution of pixels belonging to the VERTICO cluster galaxy sample extend to slightly lower molecular gas surface and SFR surface densities. There is an excess in cluster pixels above $\Sigma_\star \sim 10^8 ~ \mathrm{M_\odot~kpc^{-2}}$ with respect to the field sample, although the shape of the distributions is similar, which may be driven by an excess in central stellar features such as bars and bulges in VERTICO. The combination of lower SFR and high stellar surface densities results in the significant skew of the VERTICO pixels to lower sSFRs. The next section explores whether differences between the field and cluster samples are the result of environmental processes or internal secular evolution or a mix of both.

\section{Results}
\label{sec:results}

\begin{figure*}
    \centering
    \includegraphics{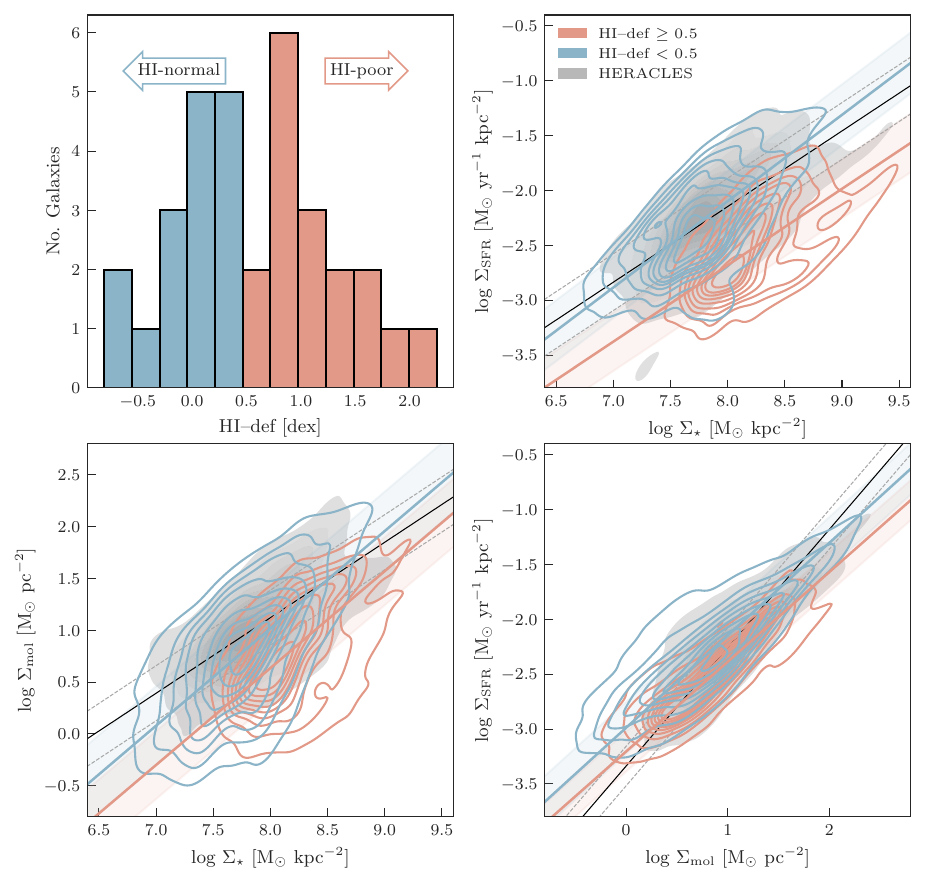}
    \caption{The effect of environment -- as traced by \HI~deficiency -- on the resolved star-forming main sequence (rSFMS, top right), molecular gas main sequence (rMGMS, bottom left), Kennicut-Schmidt relation (rKSR, bottom right) for the 15,401 pixels in our sample of 33 cluster galaxies. The same relations are shown for the resolution-matched HERACLES field sample (gray, 10,817 pixels, 10 galaxies). The distribution of \HI~ deficiencies (top left) demonstrates the selection of \HI-poor (\HI--def $\geq 0.5$ dex, red) and \HI-normal (\HI--def $< 0.5$ dex, blue) subsamples. The relationships are shown using a Gaussian kernel density estimator with ten iso-density contour levels linearly spaced between 5\% and 100\% of the distribution. The best fits are shown by the solid lines of corresponding color with the $1\sigma$ scatter illustrated by the shaded regions for the cluster subsamples and dashed lines for the field sample. Parameters and scatter of the best-fit relations are provided in Table \ref{tab:fit_params} in Appendix \ref{app:fitparameters}. The rSFMS and rMGMS in \HI-normal galaxies follow the field relationships while \HI-poor galaxies have reduced SFR and molecular gas densities at fixed stellar mass.
    \label{fig:sequences_HI_norm_poor}}
\end{figure*}

In this section, we present the VERTICO rSFMS -- the first such characterization of this relationship for cluster galaxies -- and study the VERTICO rMGMS and rKSR first published in \citet{Watts2023} and \citet{Jimenez-Donaire2022}, respectively. 

\HI~deficiency values for the VERTICO cluster galaxies are adopted from \citep{Chung2009} who compare their \HI~content with field galaxies of the same optical size, independent of morphological type. We divide the VERTICO sample described in Section \ref{sec:data} into two subsamples based on global galaxy \HI~deficiency; the \HI-poor subsample contains pixels belonging to galaxies with \HI--def $\geq 0.5$ ($6448$ pixels, 16 galaxies) while the \HI-normal subsample contains pixels drawn from galaxies with \HI--def $< 0.5$ ($8953$ pixels, 17 galaxies). The upper left panel in Figure \ref{fig:sequences_HI_norm_poor} shows this selection in the distribution of \HI~deficiencies for the 33 VERTICO galaxies studied in this work.

\subsection{The VERTICO rSFMS, rMGMS, and rKSR}

The rSFMS, rMGMS, and rKSR for the VERTICO \HI-normal (blue) and \HI-poor (red) subsamples as well as the HERACLES field sample (gray) are shown in the top left, bottom left, and bottom right panels of Figure \ref{fig:sequences_HI_norm_poor}, respectively. The corresponding best-fit relations for the \HI-normal and \HI-poor subsamples are shown by the solid lines in each plot. The fits are calculated using the {\sc LTSFIT PYTHON} package described in \citet[][Section 3.2]{Cappellari2013} which accounts for measurement uncertainties in both axes to determine the best-fit parameters and scatter. The fit parameters for each relation are provided in Appendix \ref{app:fitparameters}.

The rSFMS (top right) and rMGMS (bottom left) in \HI-normal cluster galaxies closely follow the field relations although there is a small deviation to lower gas densities at low stellar mass density in the rMGMS. This shows that where there is little effect on the \HI~content, the influence of environment on the molecular gas and subsequent star formation activity is also minimal. The rMGMS offset at the lowest stellar densities may be an indicator of environmental effects beginning to take hold in the galaxy outskirts and we return to this point with Figure \ref{fig:delta_quants_radius_Y17}.  On the other hand, \HI-poor galaxies are clearly offset to lower molecular gas and SFR densities at fixed stellar density with respect to the field relations. Thus, in galaxies where global \HI~content is reduced with respect to the field, there is also a significant reduction in molecular densities and, consequently, star formation activity.

The best-fit rSFMS relation for \HI-poor galaxies is 0.62 dex below the \HI-normal relation demonstrating that the processes that are reducing the global \HI~content are also responsible for quenching star formation within the truncation radius \citep{Watts2023} and on the scale of individual pixels ($\sim 720$~pc). 

The rMGMS fit for the \HI-poor galaxies is offset below the \HI-normal relation by 0.38 dex. Interestingly, the slopes of the \HI-poor and \HI-normal rMGMS agree well ($0.94\pm0.013$ and $0.94\pm0.010$) with the \HI-normal relation but is steeper than the field relation ($0.79\pm0.011$). One reason for this is the lack of molecular gas at low stellar surface densities ($\Sigma_\star \lesssim 10^{7.5}~\mathrm{M_\odot~kpc^{-2}}$) in \HI-poor galaxies, which implies that the impact of environment on the molecular gas disc is greatest in the low gas and stellar density regimes. In the context of the rSFMS analysis, the difference in molecular gas content for \HI-normal and -poor subsamples suggests that the link between global \HI~deficiency and subkiloparsec quenching is at least in part driven by a reduction in molecular gas density.

The \HI-poor rKSR (bottom right) is, on average, 0.24 dex below the fit to the \HI-normal galaxies. This lower SFE in regions belonging to \HI-deficient galaxies is also shown by both \citet{Villanueva2022} and \citet{Jimenez-Donaire2022}. The fact that the combined subsample offsets in rMGMS and rKSR match the observed offset in rSFMS is reflective of the fact that the location of regions in the rSFMS plane is fundamentally a consequence of their position with respect to rMGMS and rKSR \citep[e.g.,][]{Lin2019, Ellison2020, Baker2022} and this, consequently, holds true for the best-fit relations as well.

The rKSR slopes for the \HI-normal and \HI-poor subsamples are very similar ($0.84\pm0.006$ vs. $0.82\pm0.007$), but flatter than is observed for the entire sample \citep[$0.97\pm0.07$;][]{Jimenez-Donaire2022}. This is likely a consequence of a non-linear relationship between \HI~deficiency and molecular gas density meaning the offset to lower SFR density is greatest for regions with high gas density in \HI-poor galaxies. 

Although it is clear that a linear fit is not the best descriptor of the data in each relationship, our goal here is simply to provide a straightforward quantification of the offset between the \HI-poor and \HI-normal populations.

To further quantify the observed differences between regions belonging to \HI-normal and \HI-poor galaxies, we calculate the offset of each pixel from the median rSFMS, rMGMS, and rKSR of the full sample. These offset quantities are termed $\Delta \Sigma_\mathrm{SFR}$, $\Delta f_\mathrm{mol}$, and $\Delta SFE$, respectively.

We define $\Delta \Sigma_\mathrm{SFR}$ for each pixel following \citet{Ellison2020}, using relevant (VERTICO or HERACLES) sample itself to select a set of star-forming control pixels ($sSFR \geq 10^{-10.5} ~\mathrm{yr}^{-1}$) that are matched within a narrow range of stellar surface density ($\pm0.1$ dex) to the target pixel. We then calculate $\Delta \Sigma_\mathrm{SFR} = \Sigma_{\text{SFR, pixel}} - \widetilde{\Sigma}_{\text{SFR, control}}(\Sigma_\mathrm{\star})$, where $\Sigma_{\text{SFR, pixel}}$ is the SFR surface density of the pixel and $\widetilde{\Sigma}_{\text{SFR, control}}(\Sigma_\mathrm{\star})$ is the median SFR surface density of the control sample. The sSFR threshold value was chosen to maximize the difference between the pixel sSFR distributions at fixed stellar density, i.e., regions with $\Sigma_\star < 10^{8}~\mathrm{M_\odot~kpc^{-2}}$ tend to have $sSFR > 10^{-10.5} ~\text{yr}^{-1}$ and vice versa. 

The result is that the median rSFMS is defined as $\Delta \Sigma_\mathrm{SFR} = 0$, with positive and negative $\Delta \Sigma_\mathrm{SFR}$ values for regions that are more star-forming or more quenched with respect to this reference point. The normalization of this relation means that the median value of $\Delta \Sigma_\mathrm{SFR}$ for all VERTICO pixels is $-0.23$ dex or, in other words, the median VERTICO pixel is $\sim 40\%$ less star-forming than the median rSFMS.

We also quantify the molecular gas content and SFE by calculating the pixel offsets from the rMGMS and rKSR ($\Delta f_\mathrm{mol}$ and $\Delta SFE$, respectively). Briefly, $\Delta f_\mathrm{mol} = \Sigma_{\text{mol, pixel}} - \widetilde{\Sigma}_{\text{mol, control}}(\Sigma_\mathrm{\star})$, where $\Sigma_{\text{mol, pixel}}$ is the gas surface density of the pixel and $\widetilde{\Sigma}_{\text{mol, control}}(\Sigma_\mathrm{\star})$ is the median gas surface density of all pixels within $\pm0.1$ dex in stellar surface density. Similarly, $\Delta SFE = \Sigma_{\text{SFR, pixel}} - \widetilde{\Sigma}_{\text{SFR, control}}(\Sigma_\mathrm{mol})$, where $\Sigma_{\text{SFR, pixel}}$ is the SFR surface density of the pixel and $\widetilde{\Sigma}_{\text{SFR, control}}(\Sigma_\mathrm{mol})$ is the median SFR surface density of all pixels within $\pm0.1$ dex in molecular gas surface density. The only difference from the calculation of $\Delta \Sigma_\mathrm{SFR}$ is that we do not impose an sSFR cut on the control samples for $\Delta SFE$ and $\Delta f_\mathrm{mol}$. Thus, regions with positive (negative) $\Delta f_\mathrm{mol}$ values are gas normal (poor) with respect to the median rMGMS, and regions where $\Delta SFE$ is positive (negative) are forming stars more (less) efficiently than the median rKSR.

For regions plotted in Figure \ref{fig:sequences_HI_norm_poor}, the median $\Delta \Sigma_\mathrm{SFR}$ values for the \HI-normal and \HI-poor subsamples are 0 and $-0.54$ dex, respectively. This tells us, perhaps unsurprisingly, that our \HI-normal sample tends to follow the median rSFMS for the entire population. Of course, this is mainly by construction given that we only include star-forming pixels when defining the $\Delta \Sigma_\mathrm{SFR}$ control samples. Interestingly, pixels in the \HI-poor subsample are a factor of $\sim 3.5$ less star-forming than the rSFMS. 

We take this analysis further by using the rMGMS to compare $\Delta f_\mathrm{mol}$ values for our two subsamples. The median $\Delta f_\mathrm{mol}$ value for pixels in \HI-normal galaxies is 0.01 dex. On the other hand, we find that pixels belonging to the \HI-poor subsample have less molecular gas, with an median $\Delta f_\mathrm{mol}$ of $-0.39$ dex. The fact that the difference in median $\Delta \Sigma_\mathrm{SFR}$ between the \HI-normal and -poor galaxies is significantly larger than for $\Delta f_\mathrm{mol}$ tells us that suppression of molecular gas content is not the whole picture.

The regions belonging to the two subsamples also bracket the median rKSR, with pixels belonging to \HI-normal and \HI-poor galaxies having median $\Delta SFE$ values of $0$ and $-0.22$ dex, respectively. Again, this clearly shows that by selecting pixels based upon their host galaxy's \HI~deficiency, we are selecting subsamples of pixels with significantly different SFEs.  Although trivial, it is important to note that the combined difference in the mean values of $\Delta f_\mathrm{mol}$ and $\Delta SFE$ values ($0.6$ dex) between the \HI-normal and -poor subsamples closely matches the difference in $\Delta \Sigma_\mathrm{SFR}$ of $0.54$ dex. In fact, the residual $0.06$ dex is well within the typical pixel uncertainties of these quantities of 0.1 dex.

\begin{figure}
    \centering
    \includegraphics{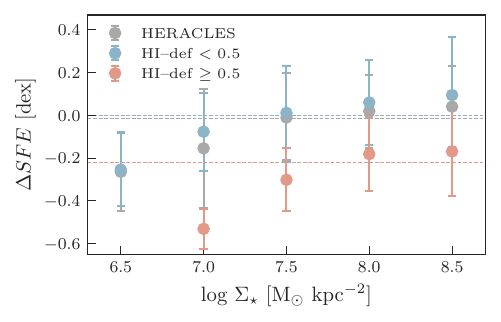}
    \caption{Median $\Delta SFE$ as a function of stellar mass surface density, $\Sigma_\star$ for the \HI-normal (blue) and \HI-poor (red) VERTICO cluster subsamples, and HERACLES field sample (gray). Error bars illustrate $\pm 1\sigma$ in each bin. The median $\Delta SFE$ (unbinned) for each subsample is shown by the dashed line of corresponding color. Pixels belonging to \HI-poor galaxies have lower SFE at fixed stellar mass surface density.}
    \label{fig:logms_delta_sfe_hi-status}
\end{figure}

It is clear from Figure \ref{fig:sequences_HI_norm_poor} that the \HI-poor galaxies do not cover the same range in stellar mass surface density as either HERACLES or \HI-normal galaxies in VERTICO. While this difference is controlled for when calculating $\Delta f_\mathrm{mol}$ and $\Delta \Sigma_\mathrm{SFR}$, it is not when computing $\Delta SFE$. Thus, it is important to check that the observed differences in the rKSR for \HI-normal and \HI-poor galaxies are not caused by stellar surface density selection effects. Figure \ref{fig:logms_delta_sfe_hi-status} shows the median $\Delta SFE$ values of pixels belonging to \HI-normal, \HI-poor, and HERACLES galaxies at fixed stellar mass surface density. In each bin, the median $\Delta SFE$ for regions in \HI-poor galaxies is lower than regions in \HI-normal and field galaxies. Interestingly, the difference between \HI-normal and \HI-poor is largest at lower $\Sigma_\star$, where one would expect environmental mechanisms to be most effective. Figure \ref{fig:logms_delta_sfe_hi-status} demonstrates that the offset of the \HI-poor rKSR to lower SFEs shown in Figure \ref{fig:sequences_HI_norm_poor} is not caused by different stellar mass density distributions between the two samples.

In an effort to understand the physics driving these trends, we now look to confirm the connection between environmental mechanisms and \HI~deficiency in our sample and connect the influence of these mechanisms on the gas-star formation cycle back to the rSFMS, rMGMS, and rKSR.

\subsection{The connection between \HI~deficiency and evolutionary stage within the cluster}

\begin{figure}
    \centering
    \includegraphics{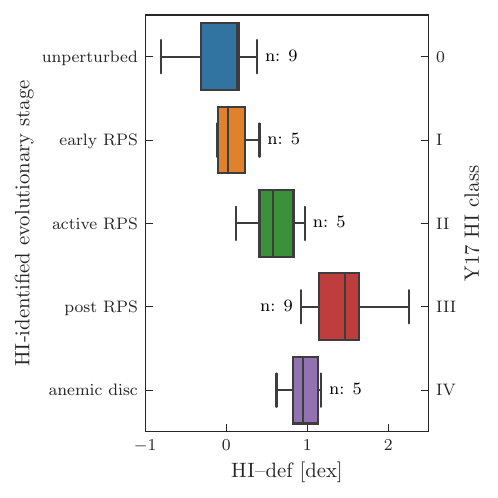}
    \caption{The distribution of galaxy \HI~deficiency for distinct ram pressure stripping stages compiled from \citet{Yoon2017}. The box shows the quartiles of the distribution while the whiskers extend to the full range in values. The number of galaxies in each stage is written next to the box. \HI~deficiency increases as ram pressure stripping proceeds.}
    \label{fig:hi-def_Y17}
\end{figure}

We interpret the differences seen in Figure \ref{fig:sequences_HI_norm_poor} as evidence that environmental processes are responsible for the systematic quenching Virgo galaxies on subkiloparsec scales. However, this assumes that \HI~deficiency is an effective quantitative measure of environmental influence on VERTICO galaxies. The large body of work that underpins this assumption is detailed in the introduction and we can also demonstrate this explicitly for our sample.

\citet{Yoon2017} provide classifications of environmental influence for the VERTICO targets based on a comparison of \HI~morphology with their orbital history. These authors identify distinct ram-pressure stripping stages (early, active, and post), as well as two other classes for unperturbed and anemic (i.e., low surface density) \HI~ discs. \citet{Yoon2017} suggest that the anemic galaxies are likely undergoing starvation and/or thermal evaporation. While this may be true, our view is that making this distinction implies that ram pressure is not at play which is unlikely since these galaxies have been in the cluster for more than one pericentre passage and have comparable \HI~extent to the other classes at fixed \HI~deficiency. Additionally, unlike the other classes, the anemic discs are also only present at high stellar mass \citep{Yoon2017}.
Thus, we suggest that anemic discs are a case where environmental processes such as ram pressure stripping have removed the \HI~ in the outskirts, but not all the \HI~ in the inner regions (perhaps due the large stellar mass of the galaxy), and there is no significant accretion of new gas onto to the disc.

Figure \ref{fig:hi-def_Y17} uses the data of \citet{Yoon2017} to illustrate the dependence of \HI~deficiency upon the gas stripping histories of galaxies previously identified by those authors. The \HI~content of galaxies is systematically depleted as galaxies transition from unperturbed discs through the early, active, and post ram pressure stripping stages. The anemic discs are also very \HI~deficient, although generally not to the extent of the highly truncated post ram pressure stripping class.

\subsection{Radial variations in environmental influence}

\begin{figure*}
    \centering
    \includegraphics{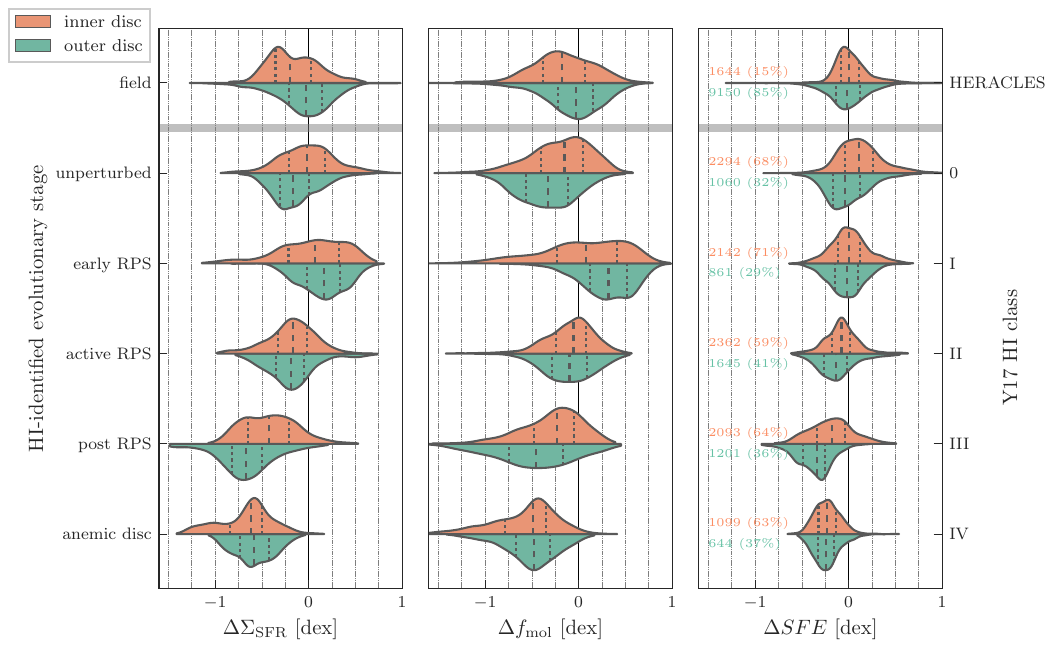}
    \caption{The distribution of $\Delta \Sigma_\mathrm{SFR}$, $\Delta f_\mathrm{mol}$, and $\Delta SFE$ pixel values for galaxies in different evolutionary stages. Field galaxies are drawn from the HERACLES sample, while \HI-identified ram pressure stripping stages and anemic discs are taken from \citet{Yoon2017}. Pixel distributions are divided into the inner and outer disc ($r < r_\mathrm{90,\,mol}$ and $r \geq r_\mathrm{90,\,mol}$, respectively). We annotate the number and percentages of pixels in the inner and outer discs for each stage in the corresponding color. The gray horizontal bar simply illustrates the distinction between our field and cluster samples. We see the systematic decrease in $\Delta \Sigma_\mathrm{SFR}$, $\Delta f_\mathrm{mol}$, and $\Delta SFE$ during the stripping sequence and also in anemic discs with respect to the field. The reduction in gas content and SFE, and consequently star formation activity, happens preferentially in the outer disks for galaxies undergoing ram pressure stripping.}
    \label{fig:delta_quants_radius_Y17}
\end{figure*}

In Figure \ref{fig:delta_quants_radius_Y17}, the panels from left to right show the distribution of $\Delta \Sigma_\mathrm{SFR}$, $\Delta f_\mathrm{mol}$, and $\Delta SFE$ pixel values as a function of host galaxy \HI-identified evolutionary stage. We include the HERACLES galaxies as our field sample in addition to the environmental classes identified by \citet{Yoon2017} and shown in Figure \ref{fig:hi-def_Y17}. The distributions are separated into pixels from the inner and outer disc based up on whether they fall inside or outside of the galaxy's 90\% molecular gas mass radius published by \citet[][$r_\mathrm{90,\,mol}$ in Table 3]{Brown2021}. The median value of each distribution is shown by the dashed lines, while the dotted lines show the 25th and 75th percentiles, respectively.

Local star formation rate decreases as a function of ram pressure stripping stage. This quenching is most apparent in the outer disc where the median values of $\Delta \Sigma_\mathrm{SFR}$ in the early, active, and post ram pressure stripping phases are 0.18, -0.18, -0.65 dex, respectively. In comparison, in the inner disc during these same phases, the median values of $\Delta \Sigma_\mathrm{SFR}$ are 0.08, -0.07 dex, and -0.12 dex. The observed star formation quenching as galaxies are stripped is primarily driven by decreases in the molecular gas content, with a secondary dependence on SFE. As with star formation activity, the effects of ram pressure stripping on molecular gas are most pronounced in the outer discs where galaxies in the early, active, and post ram pressure stripping stages have median values of $\Delta f_\mathrm{mol}$ of 0.31, -0.11, and -0.47 dex, respectively (inner disc median $\Delta f_\mathrm{mol}$ = 0.07, -0.06, -0.24 dex). There is also a significant decrease in the SFE of the outer discs during ram pressure stripping, although considerably less than for molecular gas. The median values of $\Delta SFE$ in the outer discs of early, active, and post ram pressure stripping phases are -0.01, -0.12, and -0.33 dex, respectively (inner disc median $\Delta SFE$ = 0.01, -0.07, -0.17 dex). In summary, by the time the \HI~disc of galaxy is fully stripped (`post RPS') its molecular gas content has reduced by a factor of 1.7 (0.24 dex)  and 3 (0.47 dex) in the inner and outer regions, respectively. This drives quenching in both radial bins that is compounded by the simultaneous reduction in SFE (a factor of 1.5 and 2.1, respectively).

The star formation activity of anemic discs is quenched to a similar extent to the post ram pressure stripping phase. However, unlike galaxies in the ram pressure stripping phases, anemic galaxies are quenched across the disc with a median $\Delta \Sigma_\mathrm{SFR}$ of -0.6 and -0.56 dex in the inner and outer regions, respectively. This is caused by the suppression of molecular gas content and SFE throughout the galaxy (median $\Delta f_\mathrm{mol}$ = -0.50 and -0.49 dex and median $\Delta SFE$ = -0.22 and -0.22 dex, for inner and outer discs, respectively).

Figure \ref{fig:delta_quants_radius_Y17} shows that for galaxies following the classic ram pressure stripping sequence, environment acts to decrease gas content and SFE, especially in the outskirts, resulting in the outside-in quenching of star formation. On the other hand, there is a comparable decrease in gas content and SFE for the anemic discs of high mass galaxies, but the effect is found in both the inner and outer the disc.

Even in galaxies with morphologically unperturbed \HI-discs, we see suppression of local star formation in the outer discs (median $\Delta \Sigma_\mathrm{SFR} = -0.15~$dex). This may be the result of `pre-processing' of galaxies in the group environment or outermost regions of the cluster. This appears to be driven soley by reduced gas content (median $\Delta f_\mathrm{mol} = -0.34~$dex), with SFE typical of the field (median $\Delta SFE = -0.02~$dex).

We also see enhanced star formation activity driven by increased gas densities with respect to the field in the earliest ram pressure stripping stage, with the effect again most prominent in the outer disc. The median $\Delta f_\mathrm{mol}$ in the outer discs of these galaxies is 0.31 dex, while the inner disc exhibits a typical field value of 0.07 dex. Although the number of galaxies in the stage is small (4), this result agrees with a picture from other surveys where ram pressure initially {\em increases} gas content before quenching \citep[e.g.,][]{Cramer2020, Moretti2020a, Roberts2022, Troncoso-Iribarren2020}. At face value we appear to contradict the result of \citet{Watts2023} who find no evidence for environment-driven increases in molecular gas densities in VERTICO galaxies. However, \citet{Watts2023} are comparing much coarser subsets of the VERTICO galaxies divided into three bins of \HI-deficiency. Figure \ref{fig:delta_quants_radius_Y17} demonstrates that the increase in gas densities due to ram pressure is only apparent in the earliest stage of stripping identified by the \citet{Yoon2017} morphology and phase-space analysis. Furthermore, the effect in the inner regions is relatively subtle, only becoming prominent in the outskirts of the galaxy gas disc which are not explicitly examined by \citet{Watts2023}. Our result is also in agreement with \citet{Roberts2023} who observe increased gas densities on the leading-half (with respect to their trailing-half) of VERTICO galaxies undergoing ram pressure stripping. Finally, we also see that in the field galaxies, it is the inner disc that has reduced molecular gas content and star formation activity which may be indicative of secular processes regulating the star formation cycle in these galaxies.

This analysis suggests that the environmental mechanisms causing global \HI~deficiency are also systemically and significantly reducing molecular gas content and SFE, and consequently star formation activity, on subkiloparsec scales in cluster galaxies. The influence of ram pressure stripping increases as the stripping processes proceeds, occurring preferentially in the disc outskirts. The star formation cycle of galaxies with anemic \HI~discs, likely caused by starvation of the gas supply, is impacted to a similar extent as the post-ram pressure stripping galaxies. However, the effect of starvation is commensurate in both the inner and outer disc. In the next section, we discuss the physical causes of \HI~deficiency in Virgo Cluster galaxies, their connection to molecular gas content, and the local and global star formation cycles.

\section{Discussion}
\label{sec:discussion}
Our results show that regions belonging to more \HI-deficient galaxies typically have lower molecular gas content and SFEs at fixed stellar surface density, and therefore less star formation activity than their counterparts in \HI-normal galaxies (Fig. \ref{fig:sequences_HI_norm_poor} and Fig. \ref{fig:logms_delta_sfe_hi-status}).

Figure \ref{fig:hi-def_Y17} shows that the \HI-poor subsample (\HI--def$~\geq 0.5$) includes galaxies with \HI~discs that have already been severely stripped (`post RPS'), a subset of those that are still undergoing stripping (`active RPS'), and galaxies that are primarily undergoing starvation rather than stripping (`anemic disc'). On the other hand, the \HI-normal subsample (\HI--def$~< 0.5$) combines galaxies with relatively unperturbed \HI~discs and those in the earliest phase ram pressure stripping (`early RPS') with the remaining, more \HI-rich, galaxies that are experiencing stripping (`active RPS'). Thus, the observed differences between \HI-normal and \HI-poor subsamples in the rMGMS and rKSR (Fig. \ref{fig:sequences_HI_norm_poor}) are evidence that environmental processes are driving a systematic reduction in molecular gas density and SFE on subkiloparsec scales in our galaxies. 

Figure \ref{fig:delta_quants_radius_Y17} takes the analysis further by demonstrating that for galaxies undergoing ram pressure stripping, the magnitude of this decrease is correlated with infall stage and the location within the disc. The lower median $\Delta \Sigma_\mathrm{SFR}$, $\Delta f_\mathrm{mol}$, and $\Delta SFE$ in the outer disc of active and post ram pressure stripped galaxies can be interpreted as differences in the rSFMS, rMGMS, and rKSR for regions inside and outside of the galaxy truncation radius. 

In comparison, the anemic discs have similarly effective at reducing gas content and SFE, however, this influence is seen in both the inner regions and outskirts of disc. This suggests that the gas reservoir that remains after stripping in these galaxies is primarily being consumed by ongoing star formation rather than being directly affected by external processes which one would expect to act more strongly on the outer disc.

Gas content and SFE are both reduced in galaxies undergoing ram pressure stripping and starvation, and the quantitative influence of the environmental mechanisms is always largest on the gas content. The fact that we see a significant reduction in the gas content and star formation activity in active and post RPS galaxies suggests that a significant impact on the star formation cycle only occurs in the more advanced stripping stages. In other words, there is a threshold point at which gas stripping becomes an effective quenching mechanism. Prior to this point, in the early RPS stage, we see enhanced star formation activity that is driven only by increased gas density (at fixed stellar density) with no measurable change in SFE. We also see tentative evidence for a reduction in molecular gas densities in the `unperturbed` cluster galaxies which is possibly due to pre-processing prior to cluster infall.

It is reasonable to question whether the observed trends in Figure \ref{fig:sequences_HI_norm_poor} are being wrongly attributed to environment and are instead governed by secular processes. Indeed, a number of works have shown that gas content and SFE is reduced in field galaxies transitioning from star-forming to quiescence (i.e., green-valley) using both global and resolved measurements \citep{Brownson2020, Colombo2020, Ellison2021, Lin2022, Saintonge2022}. However, the fact that the reduction in gas content and SFE is correlated with \HI-identified environmental influence {\em and} location within the galaxy disc (Fig. \ref{fig:delta_quants_radius_Y17}) suggests outside-in quenching mechanisms rather than secular processes that are associated with the presence of stellar structure in the inner disc (e.g., bars, bulges). Furthermore, there are only four active galactic nuclei (AGN) in our sample, two of which are in the \HI-poor sample, so AGN feedback is not a major driver of our results \citep[NGC 4388, NGC 4536, NGC 4579, NGC 4772;][]{Gavazzi2018, Esparza-Arredondo2018, Lee2022}.


\subsection{The VERTICO view of quenching in Virgo galaxies}

To refocus this discussion on the bigger picture of how environment regulates the star formation cycle in VERTICO galaxies, we now place our results in the context of the four previous VERTICO science papers described in the introduction; \citet[][Z22]{Zabel2022}, \citet[][V22]{Villanueva2022}, \citet[][JD22]{Jimenez-Donaire2022}, and  \citet[W23]{Watts2023}. 

Z22, V22, and W23 all find evidence for the systematic decrease in molecular gas content in \HI~deficient galaxies that we see in this work. W23 shows that this environment-driven suppression occurs across the galaxy disc, both inside and outside of the truncation radius. Along with Z22, they also demonstrate that low density molecular gas is preferentially affected. While gas content is suppressed across the disc, Z22 and V22 use radial analyses to show that molecular gas discs in VERTICO are more centrally concentrated than in field galaxies, with galaxies impacted by their environment showing steeper and/or truncated gas density profiles. These results clearly align with the offset of the \HI-poor rMGMS to lower gas densities (Fig. \ref{fig:sequences_HI_norm_poor}) and the reduction of gas content as function of environmental and location within the disc (Fig. \ref{fig:delta_quants_radius_Y17}). Taken together, this demonstrates that environmental processes have a predominantly destructive effect on molecular gas reservoirs throughout the disc, with the low-density gas in the outskirts of galaxies affected first and foremost.

As shown in this work, environmental impact on the amount of molecular gas available for star formation is not the only factor. V22 show that the SFE of molecular gas within the stellar radius decreases with increasing environmental perturbation. JD22 complete a more detailed investigation of the rKSR than in this work and are the first to report the lower SFEs in \HI-deficient galaxies with respect to \HI-normal cluster galaxies, with the latter following the relationship found in field galaxies. As well as the longer depletion times in galaxies undergoing environmental transformation, JD22 also find significant galaxy-to-galaxy and region-to-region variation in SFE, suggesting there is no single rKSR that can be used to accurately describe all cluster galaxies. We do note that this galaxy-to-galaxy diversity in SFE is also found in field galaxies \citep[e.g.,][]{Ellison2021, Casasola2022, Saintonge2022}. So, while the correlation with stripping stage and lower SFE in the outer discs in Figure \ref{fig:delta_quants_radius_Y17} suggests environmental processes are at play, we have not yet precisely determined the extent to which the cluster environment is responsible for decreasing the efficiency with which galaxies convert their molecular gas into stars.


We do, however, demonstrate that environment generally acts to reduce both molecular gas content and SFE throughout the disc of Virgo galaxies, albeit preferentially in the outer regions for galaxies undergoing stripping. In an effort to identify the particular environmental process(es) responsible for these effects, we use existing \HI~morphological classifications to show that the observed reduction in gas content, SFE, and activity in \HI-poor VERTICO galaxies increases with ram pressure stripping stage. We also investigate the effect that the lack of gas accretion after parts of the disc have been stripped (i.e., starvation) has on the star formation cycle in the remaining disc. We demonstrate that the gas content and SFE of starved galaxies is negatively influenced to the same degree as stripped galaxies, although the reduction is uniform throughout the remaining disc which suggests ongoing star formation is primarily responsible. 

In support of a picture where ram pressure strongly regulates the star formation cycle of most \HI-poor VERTICO galaxies, V22 show that the ratio of molecular to atomic gas ($R_\mathrm{mol}$) increases within the stellar radius with the level of \HI~truncation and/or asymmetry. In the same vein, W23  observe a decrease in molecular gas content and increases in $R_\mathrm{mol}$ that cannot be driven by changes in the ISM physical conditions alone and invoke stripping of the molecular gas to explain their results. The ram pressure stripping scenario is also entirely consistent with the preferential suppression of low density gas in the outskirts seen in our work, as well as the steep and truncated gas density profiles in \HI-deficient galaxies (Z22, W23, V22). 

In agreement with previous work focusing on local versus global quenching \citep[e.g.,][]{Bluck2020, Bluck2022}, our results and JD22 show that in Virgo the two go hand-in-hand. Indeed, JD22 report that \HI-deficient VERTICO galaxies are offset below the global star-forming main sequence for field galaxies by $0.6$ dex. In our work, we show subkiloparsec regions belonging to \HI-poor galaxies are, on average, a factor of $\sim3.5$ ($\sim 0.54$ dex) less star-forming than their counterparts in \HI-normal systems. Given that \HI~deficiency broadly traces stripping stage in our sample (Fig. \ref{fig:hi-def_Y17}), this shows that ram pressure stripping quenches star formation on both local and global scales in VERTICO galaxies. The analysis of Figure \ref{fig:delta_quants_radius_Y17} suggest that stripping acts from the outside-in to suppress gas content and SFE {\em throughout} the disc.

Lastly, we note that we find evidence for the elevation of star formation activity driven by increased molecular gas densities at fixed stellar mass in the earliest stripping stage, especially in the outskirts. In combination with \citet{Roberts2023}, who find gas compression on the leading half of ram pressure stripped VERTICO galaxies, these results suggest that although ram pressure is ultimately a global quenching process, its initial effect is to induce local star formation activity driven by systematic gas compression prior to removal. Neither this work nor \citet{Roberts2023} find evidence for increased SFE due to ram pressure stripping and, while JD22 find azimuthal variations in SFE in two VERTICO galaxies, only one of these (NGC 4654) has a clear signature of enhanced SFE on the leading edge. One such instance in the entire 49 CO detected galaxies in VERTICO supports our conclusion that this scenario is the exception rather than the rule, at least at the sensitivity and resolution of our observations.


\section{Summary \& Conclusions}
\label{sec:summary}
This paper presents the first galaxy cluster rSFMS using 15,401 subkiloparsec regions drawn from 33 Virgo members in the VERTICO survey. We also revisit the rMGMS and rKSR shown by previous VERTICO work (W23, JD22). The influence of environment on the subkiloparsec star formation cycle of Virgo galaxies is characterized by dividing the selected pixels into subsamples belonging to \HI-normal (\HI--def $< 0.5$) and \HI-poor (\HI--def $\geq 0.5$) galaxies. We then compare the differences in the gas content, SFE, and star formation activity of regions with respect to the median relation as a function of their location within the disc and host galaxy evolutionary stage. This analysis has produced a number of results which we summarize here:

\begin{enumerate}
    \item \HI-poor galaxies have lower SFR surface densities with respect to \HI-normal cluster and field galaxies at fixed stellar mass surface density. The median $\Delta \Sigma_\mathrm{SFR}$ of \HI-poor galaxies is 0.54 dex lower than for \HI-normal systems  (Fig. \ref{fig:sequences_HI_norm_poor}). 
    We conclude that the decrease in star formation activity in \HI-poor galaxies is due to the combined effects of reduced molecular gas content and SFE.
    \item The median molecular gas content (as quantified by $\Delta f_\mathrm{mol}$) of regions belonging to \HI-poor galaxies is a factor of three (0.54 dex) lower than regions in \HI-normal galaxies (Fig. \ref{fig:sequences_HI_norm_poor}) at fixed stellar mass surface density. The difference increases with decreasing molecular gas density, suggesting that the low-density gas is preferentially affected. This result supports previous work by Z22 and V22.
    \item Regions in the \HI-poor subsample have -0.22 dex lower SFE at fixed $\Sigma_\star$ than \HI-normal galaxies suggesting that environmental mechanisms are responsible for the observed decrease in SFE in Virgo (Fig. \ref{fig:sequences_HI_norm_poor} and Fig. \ref{fig:logms_delta_sfe_hi-status}), supporting previous work by JD22 (Fig. \ref{fig:sequences_HI_norm_poor}).
    \item We observe systematically enhanced SFR and molecular gas surface densities at fixed stellar density (0.18 dex and 0.31 dex, respectively) in the earliest stage of ram pressure stripping at fixed stellar surface density, while the SFE remains unchanged (Fig. \ref{fig:delta_quants_radius_Y17}). This effect is most pronounced in the outer discs of galaxies.
    \item Galaxies that have undergone starvation of their gas supply (likely in addition to ram pressure stripping) have similarly reduced star formation activity, molecular gas content, and SFE to the outskirts of truncated ram pressure stripped galaxies (Fig. \ref{fig:delta_quants_radius_Y17}). However, the influence of starvation on the star formation cycle occurs throughout the disc, suggesting that it is an effective quenching mechanism in the remaining gas disc not affected by ram pressure which primarily affects the outer regions.
\end{enumerate}

This paper presents a coherent picture of how environmental mechanisms, in particular ram pressure stripping and starvation, regulate the star formation cycle in VERTICO galaxies. Future work will focus on outstanding questions such as; what are the physical drivers of the variation in SFE? How does environment regulate molecular gas kinematics? By answering these questions, VERTICO will give us a more complete understanding of the gas star formation cycle in cluster galaxies.


\section{Data Availability}
The molecular gas, SFR and stellar mass surface density maps for all VERTICO galaxies are publicly available for download from the Canadian Advanced Network for Astronomical Research VOSpace cloud storage\footnote{\href{https://www.canfar.net/storage/vault/list/VERTICO}{https://www.canfar.net/storage/vault/list/VERTICO}}. Note that the online maps are not corrected for galaxy inclination. The data set and the Jupyter notebook used to perform this analysis are publicly available on GitHub\footnote{\href{https://github.com/drtobybrown/vertico_main_sequence}{https://github.com/drtobybrown/vertico\_main\_sequence}}.

\section{Acknowledgments}
The majority of this work was conducted on the traditional territory of the T'Sou-ke and Lekwungen peoples. The authors acknowledge and respect the T'Sou-ke, Songhees, Esquimalt and WS\'{A}NE\'{C} Nations whose historical relationships with the land continue to this day.

We thank the anonymous referee for a considered review of the manuscript.

This work was carried out as part of the VERTICO collaboration.
TB acknowledges support from the National Research Council of Canada via the Plaskett Fellowship of the Dominion Astrophysical Observatory. IDR acknowledges financial support from the ERC Starting Grant Cluster Web 804208 (P.I. van Weeren).
NZ is supported by the South African Research Chairs Initiative of the Department of Science and Technology and National Research Foundation.
CDW acknowledges support from the Natural Sciences and Engineering Research Council of Canada and the Canada Research Chairs program.
LC acknowledges support from the Australian Research Council via the Future Fellowship and Discovery Project funding schemes (FT180100066, DP210100337). Parts of this research were conducted by the Australian Research Council Centre of Excellence for All Sky Astrophysics in 3 Dimensions (ASTRO 3D), through project number CE170100013. 
AC acknowledges support by the National Research Foundation of Korea (NRF), grant Nos. 2022R1A2C100298211, and 2022R1A6A1A03053472. 
TAD acknowledges support from the UK Science and Technology Facilities Council through grant ST/W000830/1.
BL acknowledges the support from the Korea Astronomy and Space Science Institute grant funded by the Korea government (MSIT) (Project No. 2022-1-840-05).
RS acknowledges financial support from FONDECYT Regular 2023 project No. 1230441
V.V. acknowledges support from the scholarship ANID-FULBRIGHT BIO 2016-56160020 and funding from NRAO Student Observing Support (SOS)—SOSPA7-014. V.V. acknowledge partial support from the NSF grants AST2108140 and AST1615960.

Parts of this research were supported by the Australian Research Council Centre of Excellence for All Sky Astrophysics in 3 Dimensions (ASTRO 3D), through project number CE170100013.

This paper makes use of the following ALMA data: 

\begin{itemize}
    \item ADS/JAO.ALMA\href{https://almascience.nrao.edu/asax/?result_view=observation&projectCode=\%222019.1.00763.L\%22}{\#2019.1.00763.L} 
    \item ADS/JAO.ALMA\href{https://almascience.nrao.edu/asax/?result_view=observation&projectCode=\%222017.1.00886.L\%22}{\#2017.1.00886.L} 
    \item ADS/JAO.ALMA\href{https://almascience.nrao.edu/asax/?result_view=observation&projectCode=\%222016.1.00912.S\%22}{\#2016.1.00912.S} 
    \item ADS/JAO.ALMA\href{https://almascience.nrao.edu/asax/?result_view=observation&projectCode=\%222015.1.00956.S\%22}{\#2015.1.00956.S}
\end{itemize}

ALMA is a partnership of ESO (representing its member states), NSF (USA) and NINS (Japan), together with NRC (Canada), MOST and ASIAA (Taiwan), and KASI (Republic of Korea), in cooperation with the Republic of Chile. The Joint ALMA Observatory is operated by ESO, AUI/NRAO and NAOJ. The National Radio Astronomy Observatory is a facility of the National Science Foundation operated under cooperative agreement by Associated Universities, Inc.

This research made use of data provided by NASA/IPAC Infrared Science Archive, which is funded by the National Aeronautics and Space Administration and operated by the California Institute of Technology.

The authors acknowledge the use of the Canadian Advanced Network for Astronomy Research (CANFAR) Science Platform. Our work used the facilities of the Canadian Astronomy Data Center, operated by the National Research Council of Canada with the support of the Canadian Space Agency, and CANFAR, a consortium that serves the data-intensive storage, access, and processing needs of university groups and centers engaged in astronomy research \citep{Gaudet2010}.


%

\facilities{ALMA, GALEX, Sloan, WISE}


\software{\href{https://github.com/Stargrazer82301/AncillaryDataButton}{The Ancillary Data Button} \citep{Clark2018}, \href{http://www.astropy.org}{\sc Astropy} \citep{astropy:2013, astropy:2018,larry_bradley_2020_4044744}, 
\href{https://matplotlib.org/}{\sc Matplotlib} \citep{Hunter2007, 2020SciPy-NMeth}, 
\href{https://pandas.pydata.org/}{\sc Pandas} \citep{mckinney-proc-scipy-2010}, 
\href{https://github.com/johannesjmeyer/rsmf}{\sc Rsmf}, 
\href{https://scipy.org/}{SciPy} \citep{2020SciPy-NMeth}, 
\href{https://seaborn.pydata.org/index.html}{\sc Seaborn} \citep{Waskom2021}}



\clearpage

\appendix

\section{Best-fit parameters for Figure 2}
\label{app:fitparameters}

Table \ref{tab:fit_params} provides the best-fit parameters and scatter for the \HI-poor and \HI-normal subsamples in the rSFMS, rMGMS, and rKSR shown in Figure \ref{fig:sequences_HI_norm_poor}. The fits and scatter are derived using the {\sc LTSFit} package, for more details see Section 3.2 of \citet{Cappellari2013}. In this work, we keep the limit on the fraction of possible outliers at its default value of 50\% but this choice does not strongly influence our results.

\begin{deluxetable}{cccccc}[h]
\tablecaption{Best-fit parameters of the relation $y = a + (bx - \mathrm{pivot})$, 
where $x$ and $y$ are the respective axes of the relation in column (1). The units are given in Figure \ref{fig:sequences_HI_norm_poor}. Pivot is in the same units as $x$. The rms scatter, $\sigma$, of the data about the fitted relation is calculated as the standard deviation of $y_\mathrm{fit} - y_\mathrm{pix}$. For the rKSR, we fit for $\Sigma_\mathrm{mol}$ as the uncertainties are larger than for $\Sigma_\mathrm{SFR}$ and rearrange the fit result for $y = c + mx$ where $c = -a/b+\mathrm{pivot}$ and $m = 1/b$.  \label{tab:fit_params}}
\tablehead{\colhead{Relation} & \colhead{subsample} & \colhead{$a$} & \colhead{$b$} & \colhead{pivot} & \colhead{$\sigma$}}
\startdata
rSFMS & HERACLES & -2.26$\pm$0.003 & 0.69$\pm$0.006 & 7.84 & 0.26 \\
      & \HI-poor & -2.57$\pm$0.003 & 0.70$\pm$0.010 & 8.16 & 0.27 \\
      & \HI-normal & -2.29$\pm$0.003 & 0.79$\pm$0.008 & 7.76 & 0.28 \\
\hline
rMGMS & HERACLES & 1.00$\pm$0.003 & 0.73$\pm$0.008 & 7.84 & 0.27 \\
      & \HI-poor & 0.78$\pm$0.004 & 0.94$\pm$0.013 & 8.16 & 0.33 \\
      & \HI-normal & 0.79$\pm$0.004 & 0.94$\pm$0.010 & 7.76 & 0.38 \\
\hline
rKSR & HERACLES & 1.02$\pm$0.003 & 0.93$\pm$0.008 & -2.24 & 0.18 \\
     & \HI-poor & 0.75$\pm$0.002 & 1.22$\pm$0.007 & -2.59 & 0.18 \\
     & \HI-normal & 0.78$\pm$0.003 & 1.19$\pm$0.006 & -2.33 & 0.24
\enddata
\end{deluxetable}

\bibliography{refs}{}
\bibliographystyle{aasjournal}



\end{document}